\documentclass[twocolumn,pra,aps]{revtex4}
\usepackage{graphicx}

\def\duzomniejsze{<\kern-.7mm<}
\def\duzowieksze{>\kern-.7mm>}

\def\textbf#1{{\bf #1}}
\def\beq{\begin{equation}}
\def\eeq{\end{equation}}
\def\be{\begin{equation}}
\def\ee{\end{equation}}
\def\ben{\begin{eqnarray}}
\def\een{\end{eqnarray}}
\def\beqa{\begin{eqnarray}}
\def\eeqa{\end{eqnarray}}
\def\eea{\end{array}}
\def\bea{\begin{array}}
\newcommand{\bei}{\begin{itemize}}
\newcommand{\eei}{\end{itemize}}
\newcommand{\bee}{\begin{enumerate}}
\newcommand{\eee}{\end{enumerate}}

\def\>{\rangle}
\def\<{\langle}

\begin{document}

\title{Entanglement and quantum phase transition in quantum mixed spin chains}

\begin{abstract}
The ground entanglement and thermal entanglement in quantum mixed
spin chains consisting of two integer spins 1 and two half integer
spins $\frac{1}{2}$ arrayed as $\frac{1}{2}-\frac{1}{2}-1-1$ in a
unit cell with antiferromagnetic nearest-neighbor couplings
$J_1$($J_2$) between the spins of equal (different) magnitudes,
are investigated by adopting the log-negativity. The ground
entanglement transition found here is closely related with the
valence bond state transition, and the thermal entanglement near
the critical point is calculated and shown that two distinct
behaviors exist in the nearest neighbor same kind of spins and
different kind of spins, respectively. The potential application
of our results on the quantum information
processing is also discussed.\\

\end{abstract}
\author{Shang-Bin Li}\email{sbli@zju.edu.cn}, \author{Zhao-Xing Xu}, \author{Jian-Hui Dai}, \author{Jing-Bo Xu}\email{phyxujb@mail.hz.zj.cn}

\affiliation{Chinese Center of Advanced Science and Technology
(World Laboratory), P.O.Box 8730, Beijing, People's Republic of
China;} \affiliation{Zhejiang Institute of Modern Physics and
Department of Physics, Zhejiang University, Hangzhou 310027,
People's Republic of China}

\maketitle

\section * {I. INTRODUCTION}

Entanglement, one of the most striking features of quantum
mechanics \cite{Einstein1935}, has been recognized as an important
resource for quantum information processing
\cite{Cirac2000,Bennett2000}. Recently, the entanglement in the
systems of condensed matter physics such as the Heisenberg models
has cause much attention
\cite{Connor2001,Arnesen2001,Zanardi2002,Wang2001,Osterloh2002,Osborne2002,Vidal2003,Shi2003,Wang2002,Bose2002,Saguia2003,Schliemann2003,Li2003,Glaser2003,Verstraete2004,Wu2004,Jordan2004,Somma2004,Wang2005,Verstraete20041,Verstraete20042,Pachos2004,Fan2004}.
Natural entanglement in the thermal equilibrium of the spin chains
coupled by the exchange interaction has been found and it is shown
that the temperature or the external magnetic field can enhance
the pairwise entanglement of spins \cite{Arnesen2001}. The
relations between the ground state entanglement and quantum phase
transition have also be revealed for the classes of spin-1/2
chains \cite{Osterloh2002,Vidal2003,Shi2003}. All results
concerning the entanglement in the Heisenberg spin chains showed
that the entanglement can make a bridge across the condensed
matter physics and quantum information theory.

One of the most famous results in quantum spin chains is the
Haldane conjecture which implies that the half-odd-integer
antiferromagnetic Heisenberg chains should have gapless spectrum
with algebraic decay of correlations at zero temperature and
integer spin ones should be gapped with exponential decay of
correlations. Recently, some authors have addressed whether the
entanglement length is finite or infinite in the Heisenberg spin-1
chain which is a gapped quantum Hamiltonian and proven the
existence of an infinite entanglement length as opposed to their
finite correlation length \cite{Verstraete20041}. Besides the
spin-1 chains, recent experimental achievements
\cite{Verdaguer1984,Kahn1992,Zheludev2001} have stimulated the
interest to study both analytically and numerically the mixed
quantum spin chains, such as the alternating quantum spin chain
with $S^{1}=1/2$ and $S^{2}=1$ with antiferromagnetic
nearest-neighbor exchange interaction or more complicated
hypothetical 1D periodic structures of different spins with
ferrimagnetic properties \cite{Hagiwara1998}. The topology of spin
arrangements in the ferrimagnetic chain plays an essential role on
the variation of energy gap associated with Haldane conjecture and
on the magnetism of ground states. The ground state of
ferrimagnetic chain with two kinds of different spins is either a
spin singlet or ferrimagnetic due to the setup of the bipartite
lattices following Marshall theorem \cite{Auerbach1994}. By
applying the Lieb-Schultz-Mattis theorem \cite{Lieb}, one can find
whether the ferrimagnetic chains have a gapless excitation though
the theorem failed to predict the existence of a gapped
excitation. In this Letter, we consider the ground state
entanglement and thermal state entanglement in an interesting
model which is called as quantum mixed spin 1/2-1/2-1-1 chain by
making use of the quantum Monte Carlo method with a improved loop
cluster algorithm \cite{Evertz1993}. Over the past years, the
quantum Monte Carlo method with a improved loop cluster algorithm
has proven it to be a efficient numerical tool to deal with the
complicated coupled spin chains \cite{Evertz1993,Xu2003}. So it is
very credible to numerically investigate the nonclassical
properties such as the entanglement of the mixed spin chain by
adopting this method.

For first time, we investigate the ground state entanglement of
the mixed spin chain up to 128 sites by exploring the
log-negativity as an entanglement measure. Two distinct behaviors
of pairwise entanglement are found. We provide a evidence that the
ground state entanglement between two 1/2 spins or between 1/2
spin and 1 spin are closely related to the so-called VBS phase
transition. At the point of VBS phase transition, the ground state
between two 1/2 spins can be transferred to two different kinds of
spins i.e. the spin 1/2 and spin 1 or vice versa. Furthermore, the
thermal state pairwise entanglement is also studied and whether
the thermal entanglement can be improved by increasing the
temperature or not is also addressed, which strongly depends on
the parameter $\alpha=J_2/J_1$.

This paper is organized as follows: In Sec.II, we briefly outline
the basic content of the quantum mixed spin chain and investigate
the ground state entanglement between two nearest-neighbor spins
$\frac{1}{2}$ or between two nearest-neighbor spins $\frac{1}{2}$
and 1. In Sec.III, we investigate the thermal state entanglement
between two nearest-neighbor spins $\frac{1}{2}$ or between two
nearest-neighbor spins $\frac{1}{2}$ and 1. It is shown that the
entanglement can be enhanced by improve the temperature. In
Sec.IV, there are some conclusive remarks.

\section * {II. The ground state entanglement and the VBS phase transition of quantum mixed spin chain }

\begin{figure}
\centerline{\includegraphics[width=2.5in]{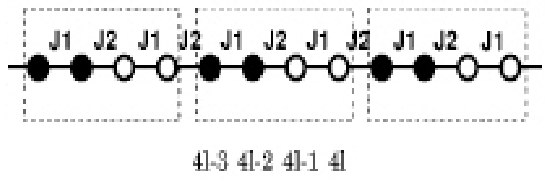}}
\caption{Graphical representation of the Hamiltonian in Eq.(1),
the black and white circles represent spins $\frac{1}{2}$ and $1$,
respectively.}
\end{figure}

Now we consider the quantum antiferromagnetic mixed spin
$S^{1}-S^{1}-S^{2}-S^{2}$ chain displayed by Fig.1, whose
Hamiltonian can be given by \cite{Xu2003} \beqa
H&=&\sum^{N/4}_{n=1}(J_1S^{1}_{4n-3}\cdot{S}^{1}_{4n-2}+J_2S^{1}_{4n-2}\cdot{S}^{2}_{4n-1}\nonumber\\
&&+J_1S^{2}_{4n-1}\cdot{S}^{2}_{4n}+J_2S^{2}_{4n}\cdot{S}^{1}_{4n+1}).
\eeqa Here, the periodic boundary condition is adopted. In the
case with $S^{1}=\frac{1}{2}$ and $S^{2}=1$, the model has been
studied by quantum Monte Carlo (QMC) simulation
\cite{Tonegawa1998} and the nonlinear $\sigma$ model (NLSM)
\cite{Fukui1997,Takano1999}. The ground state of this model is
nonmagnetic with gapped excitations. The gap varies as a function
of the parameter $\alpha=J_2/J_1$, and vanishes at a critical
point $\alpha_c$, which implies a quantum phase transition between
two different VBS state. Since the VBS states of the mixed spin
chains actually contain an ensemble of spin singlet state which
can be regards as the abundance quantum information resource. It
is very desirable to study how the parameter $\alpha$ affects the
entanglement of this model or what is the relation between the
gapped excitation and the thermal entanglement.

We study both the ground state entanglement and the thermal state
entanglement of this system. The state of a quantum system at
thermal equilibrium can be described by the density operator
$\rho=\frac{1}{Z}\exp(-\beta{H})$, where $\beta=\frac{1}{k_BT}$
with $k_B$ the Boltzmann's constant.
$Z={\mathrm{Tr}}e^{-\beta{H}}$ is the partition function. The
entanglement in the ground state or thermal state is called ground
state entanglement or thermal entanglement, respectively. For
antiferromagnetic mixed spin chains, the ground state is not
degenerate. So the ground state can be directly expressed by the
density operator $\rho=\frac{1}{Z}\exp(-\beta{H})$ with
$\beta\rightarrow\infty$. By utilizing the SU(2) invariance of
this system, the reduced density matrix $\rho^{(1,1)}$ for two
spins $\frac{1}{2}$ or $\rho^{(1,2)}$ for one spin $\frac{1}{2}$
and one spin $1$ can be obtained as follows \cite{Schliemann2003}:
\beqa
\rho^{(1,1)}=g^{(1,1)}|0,0\rangle\langle0,0|+\frac{1-g^{(1,1)}}{3}\sum^{1}_{i=-1}|1,i\rangle\langle1,i|,
\eeqa and \beqa
\rho^{(1,2)}=\frac{g^{(1,2)}}{2}\sum^{1/2}_{i=-1/2}|\frac{1}{2},i\rangle\langle\frac{1}{2},i|+\frac{1-g^{(1,2)}}{4}\sum^{3/2}_{i=-3/2}|\frac{3}{2},i\rangle\langle\frac{3}{2},i|,
\eeqa where $|J,J_z\rangle$ denotes the state of total spin $J$
and $z$ component $J_z$, and $g^{(1,1)}$ and $g^{(1,2)}$ are the
function of temperature. For characterizing the ground state
pairwise entanglement and the thermal state pairwise entanglement,
we adopt the log-negativity as the entanglement measure, which has
been proven to be an operational good entanglement measure. The
log-negativity $N(\rho_r)$ for the two particle reduced density
operator $\rho_r$ is defined as \cite{Vidal} \be
N(\rho_r)=\log_{2}\|\rho^{\Gamma}_r\|, \ee where $\rho^{\Gamma}_r$
is the partial transpose of $\rho_r$ and $\|\rho^{\Gamma}_r\|$
denotes the trace norm of $\rho^{\Gamma}_r$, which is the sum of
the singular values of $\rho^{\Gamma}_r$. For the isotropic mixed
spin chains with SU(2) symmetry, the log-negativity between any
two spins can be proven to be directly related with the
two-particle correlation function. Therefore, we can directly
calculate the log-negativity by slightly changing the program for
the QMC. In Fig.2, we numerically calculate the log-negativity
$N^{(1,1)}$ of the reduced density operator of two
nearest-neighbor spin $\frac{1}{2}$ in a unit cell of the ground
state and the log-negativity $N^{(1,2)}$ of two nearest-neighbor
different spin $\frac{1}{2}$ and $1$. It is shown that $N^{(1,1)}$
decreases from 1 to zero as the parameter $\alpha$ increases from
0 to 1. At the point of VBS phase transition labelled by
$\alpha_c\simeq{0.768}$, the entanglement experience a sharply
descent. While the entanglement between two nearest-neighbor
different spins $\frac{1}{2}$ and 1 characterized by
log-negativity $N^{(1,2)}$ is zero for small values of $\alpha$,
and near the critical point it increases with $\alpha$ and
experience a sharply ascent at the critical point. Unfortunately,
we find that the ground state entanglement only exists between two
spins in the same unit cell. In the numerical calculation of the
log-negativity of the ground state, we have set $\beta=200$ and
$N=128$. We conjecture that the descent or ascent of
log-negativity near the critical point can be more sharp if the
value of $\beta$ is chosen very very large, though this can
consume more and more computation time. From Fig.2, one can
imagine that the quantum mixed spin chain in zero temperature can
be utilized as a very good entanglement switch by controlling the
parameter $\alpha$. Recently, Pachos and Plenio have shown that
some kinds of atomic lattice can be use to simulation the
Hamiltonian of various controllable Heisenberg spin chain
\cite{Pachos2004}. It can motivate us to realize the quantum mixed
spin chain in the system of atomic lattice. Then we easily obtain
the quantum mixed spin chain with controllable parameter $\alpha$.
The details of suggestion is discussed elsewhere.

\section * {III. The thermal state entanglement of quantum mixed spin chain }

In what follows, we turn to consider the thermal entanglement of
the mixed spin chain. In Fig.3, we calculate the log-negativity
$N^{(1,1)}$ characterizing the entanglement of two
nearest-neighbor as the function of $k_BT$. It is shown that, if
the parameter $\alpha$ is smaller than the critical point
$\alpha_c$, the entanglement decreases with the temperature and
vanishes when the temperature goes beyond a threshold value.
However, in the cases with $\alpha>\alpha_c$, the entanglement
firstly increases with temperature and achieves a local maximal
value, then decreases with temperature and eventually vanishes. We
can also find that in the cases with low temperature, the
entanglement between two nearest-neighbor spin $\frac{1}{2}$
significantly decreases with $\alpha$. It is natural to conjecture
that the thermal state entanglement is closely related with fact
whether this system has the gapped excitation.

In Fig.4, we plotted the log-negativity $N^{(1,2)}$ as the
function of $k_BT$ for three different values of $\alpha$ near the
critical point. Being distinct from the entanglement of two spin
$\frac{1}{2}$, if the parameter $\alpha$ is larger than the
critical point, the entanglement between two nearest-neighbor
different spin $\frac{1}{2}$ and 1 decreases with the temperature
and vanishes when the temperature goes beyond another threshold
value which is about half of the threshold value concerning the
two spin $\frac{1}{2}$. In the case with $\alpha<\alpha_c$, the
entanglement of spin $1/2$ and spin 1 firstly increases with the
temperature, and achieves a local maximal value and then decreases
with temperature. From Fig.4, we can also find that the thermal
entanglement increases with $\alpha$, which is consistent with the
ground state entanglement.

For more deeply understanding the relation between thermal
entanglement and the gapped excitation, the analysis of the
entanglement of the excited state of the quantum mixed spin chain
is necessary. In what follows, we briefly present a simple
numerical method to calculate the log-negativity of the first
excited state. In fact, the reduced density matrix of two spins in
the first excited state can be obtained by the following
procedure. \be
\rho=\frac{1}{Z}e^{-\beta{H}}\approx\frac{1}{Z}(e^{-\beta{E_g}}|g\rangle\langle{g}|+e^{-\beta{E_1}}|e_1\rangle\langle{e_1}|)
\ee if the temperature is very very low. Then the reduced density
matrix of any two spins can also be expressed as the weight sum of
the reduced density matrices of the ground state and the first
excited state. By iterating this procedure, we can obtain the
log-negativity of two spins in any excited state. The details of
this work will be presented elsewhere.

\begin{figure}
\centerline{\includegraphics[width=2.5in]{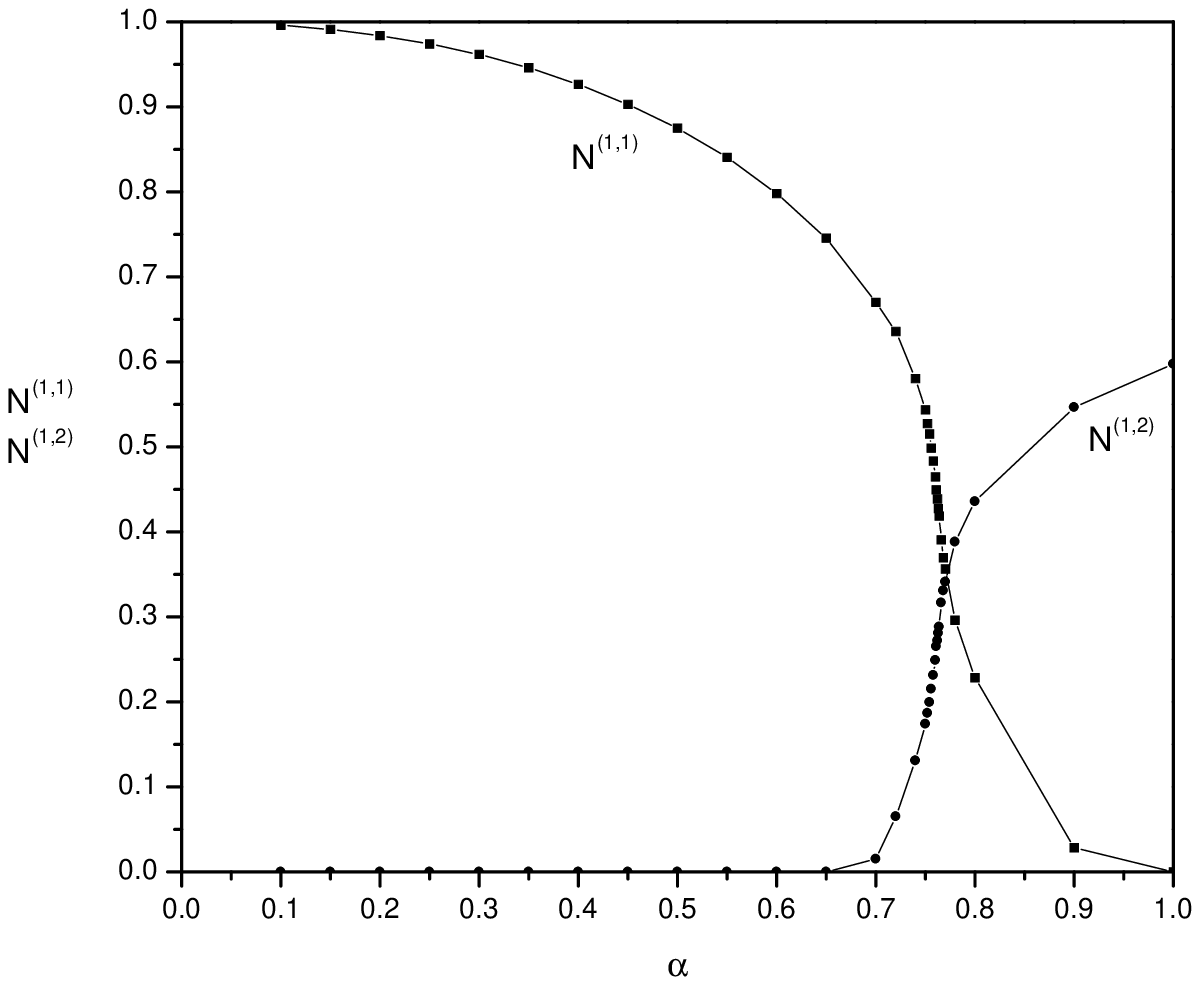}}
\caption{The log-negativity $N^{(1,1)}$ of the reduced density
operator of two nearest-neighbor spin $\frac{1}{2}$ in a unit cell
of the ground state and the log-negativity $N^{(1,2)}$ of two
nearest-neighbor different spin $\frac{1}{2}$ and $1$ are plotted
as the function of $\alpha$. The length of the lattice sites is
chosen as $N=128$ and the inverse temperature
$\beta\equiv\frac{1}{k_BT}=200$.}
\end{figure}
\begin{figure}
\centerline{\includegraphics[width=2.5in]{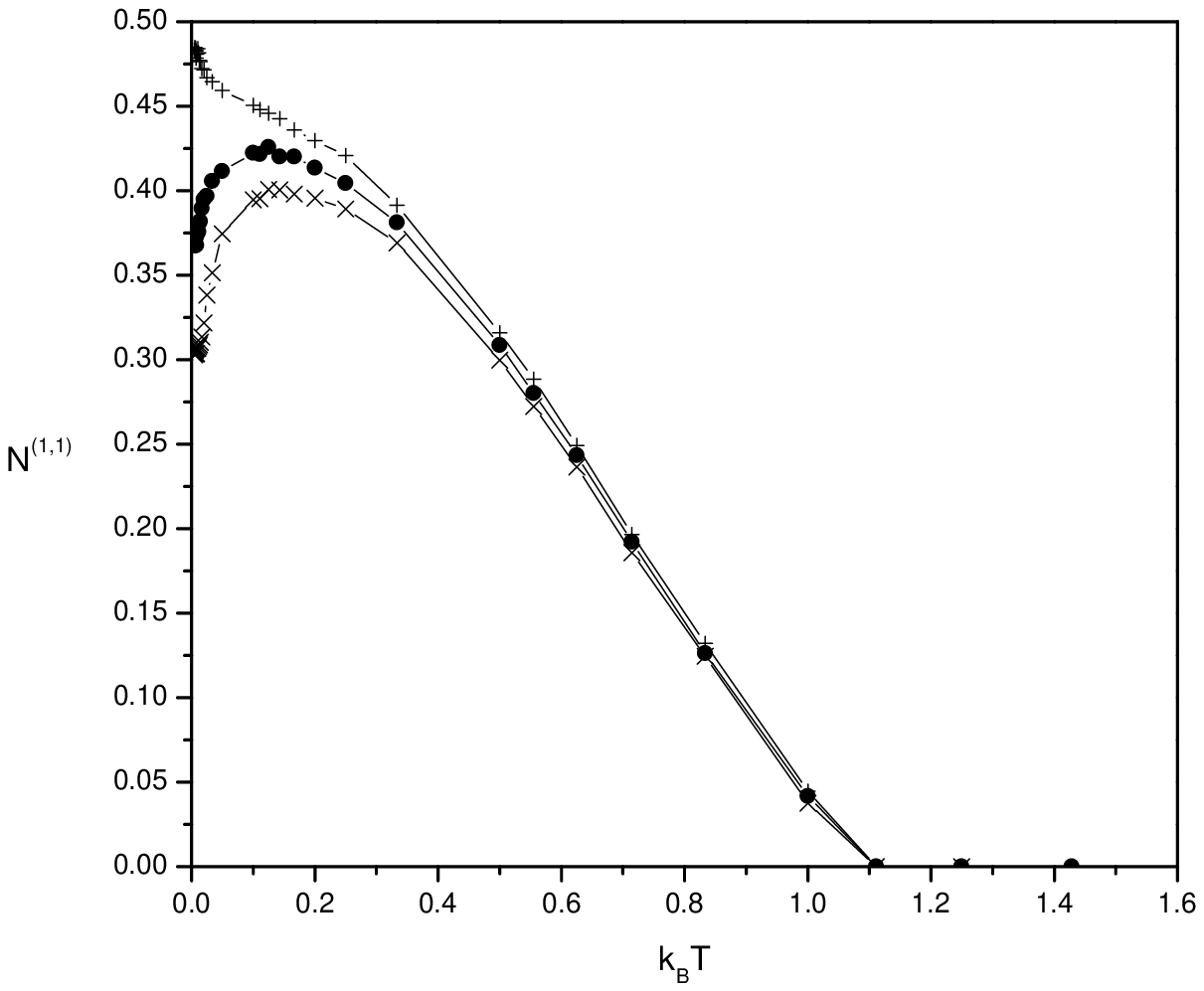}}
\caption{The log-negativity $N^{(1,1)}$ of the reduced density
operator of two nearest-neighbor spin $\frac{1}{2}$ in a unit cell
of the thermal state is plotted as the function of $k_BT$. The
length of the lattice sites is chosen as $N=128$. From top to
bottom, $\alpha=0.758, 0.768, 0.778$, respectively.}
\end{figure}
\begin{figure}
\centerline{\includegraphics[width=2.5in]{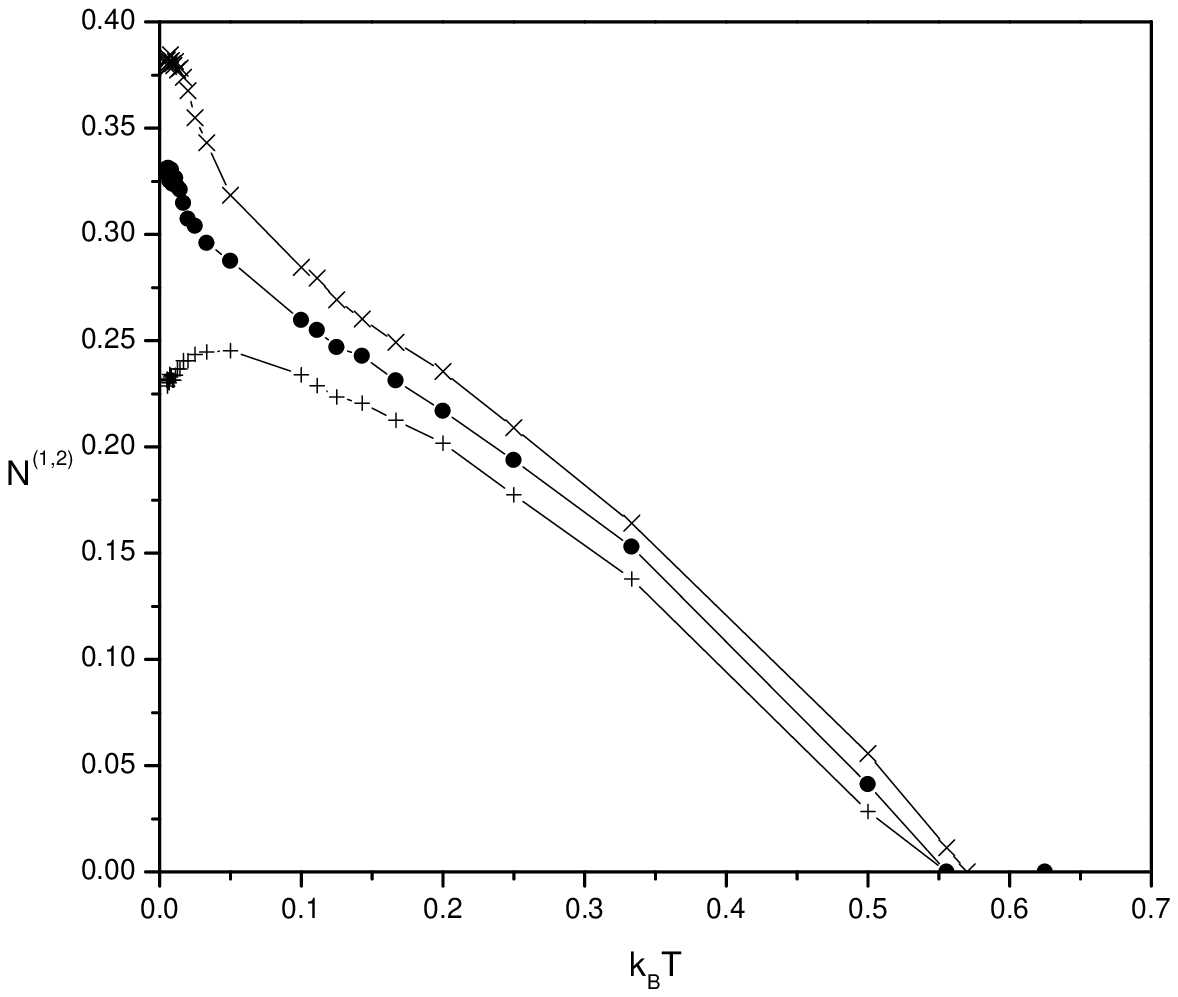}}
\caption{The log-negativity $N^{(1,2)}$ of the reduced density
operator of two nearest-neighbor spin $\frac{1}{2}$ and 1 in a
unit cell of the thermal state is plotted as the function of
$k_BT$. The length of the lattice sites is chosen as $N=128$. From
top to bottom, $\alpha=0.778, 0.768, 0.758$, respectively.}
\end{figure}

\section * {V. CONCLUSSION}

In this paper, we we investigate the ground state entanglement of
the mixed spin chain up to 128 sites by exploring the
log-negativity as an entanglement measure. Two distinct behaviors
of pairwise entanglement are found. We provide a evidence that the
ground state entanglement between two 1/2 spins or between 1/2
spin and 1 spin are closely related to the so-called VBS phase
transition. At the point of VBS phase transition, the ground state
between two 1/2 spins can be transferred to two different kinds of
spins i.e. the spin 1/2 and spin 1 or vice versa. Furthermore, the
thermal state pairwise entanglement is also studied and whether
the thermal entanglement can be improved by increasing the
temperature or not is also addressed, which strongly depends on
the parameter $\alpha=J_2/J_1$. It is shown that $N^{(1,1)}$
decreases from 1 to zero as the parameter $\alpha$ increases from
0 to 1. At the point of VBS phase transition labelled by
$\alpha_c\simeq{0.768}$, the entanglement experience a sharply
descent. While the entanglement between two nearest-neighbor
different spins $\frac{1}{2}$ and 1 characterized by
log-negativity $N^{(1,2)}$ is zero for small values of $\alpha$,
and near the critical point it increases with $\alpha$ and
experience a sharply ascent at the critical point. Unfortunately,
we find that the ground state entanglement only exists between two
spins in the same unit cell. In the numerical calculation of the
log-negativity of the ground state, we have set $\beta=200$ and
$N=128$. if the parameter $\alpha$ is smaller than the critical
point $\alpha_c$, the entanglement decreases with the temperature
and vanishes when the temperature goes beyond a threshold value.
However, in the cases with $\alpha>\alpha_c$, the entanglement
firstly increases with temperature and achieves a local maximal
value, then decreases with temperature and eventually vanishes. We
can also find that in the cases with low temperature, the
entanglement between two nearest-neighbor spin $\frac{1}{2}$
significantly decreases with $\alpha$. Being distinct from the
entanglement of two spin $\frac{1}{2}$, if the parameter $\alpha$
is larger than the critical point, the entanglement between two
nearest-neighbor different spin $\frac{1}{2}$ and 1 decreases with
the temperature and vanishes when the temperature goes beyond
another threshold value which is about half of the threshold value
concerning the two spin $\frac{1}{2}$. In the case with
$\alpha<\alpha_c$, the entanglement of spin $1/2$ and spin 1
firstly increases with the temperature, and achieves a local
maximal value and then decreases with temperature. It is natural
to conjecture that the thermal state entanglement is closely
related with fact whether this system has the gapped excitation.

\section*{ACKNOWLEDGMENT}
This project was supported by the National Natural Science
Foundation of China (Project NO. 10174066).

\bibliographystyle{apsrev}
\bibliography{refmich,refjono}

\end{document}